\documentclass[letterpaper,twocolumn,pra,
aps,showpacs,superscriptaddress,floatfix]{revtex4-2}

\usepackage[latin1]{inputenc}
\usepackage{bbm}
\usepackage{bm}
\usepackage[usenames]{color}
\usepackage{multirow}
\usepackage{amssymb}
\usepackage{amsbsy}
\usepackage{mathtools}
\usepackage{amsmath}
\usepackage{stmaryrd}
\usepackage{graphicx}
\usepackage{epsfig}
\usepackage{placeins}
\usepackage{bbold}
\usepackage{braket}
\usepackage{blindtext}
\usepackage[colorlinks,linkcolor=blue,citecolor=blue,urlcolor=blue]{hyperref}

\usepackage{scalerel}

\usepackage{tikz}
\usetikzlibrary{calc}
\usetikzlibrary{patterns}
\usetikzlibrary{svg.path}
\definecolor{orcidlogocol}{HTML}{A6CE39}
\tikzset{
  orcidlogo/.pic={
    \fill[orcidlogocol] svg{M256,128c0,70.7-57.3,128-128,128C57.3,256,0,198.7,0,128C0,57.3,57.3,0,128,0C198.7,0,256,57.3,256,128z};
    \fill[white] svg{M86.3,186.2H70.9V79.1h15.4v48.4V186.2z}
                 svg{M108.9,79.1h41.6c39.6,0,57,28.3,57,53.6c0,27.5-21.5,53.6-56.8,53.6h-41.8V79.1z M124.3,172.4h24.5c34.9,0,42.9-26.5,42.9-39.7c0-21.5-13.7-39.7-43.7-39.7h-23.7V172.4z}
                 svg{M88.7,56.8c0,5.5-4.5,10.1-10.1,10.1c-5.6,0-10.1-4.6-10.1-10.1c0-5.6,4.5-10.1,10.1-10.1C84.2,46.7,88.7,51.3,88.7,56.8z};
  }
}

\newcommand\orcid[1]{\!%
  \href{https://orcid.org/#1}{%
    \mbox{%
      \scaleto{%
        \begin{tikzpicture}[yscale=-1,transform shape]
          \pic{orcidlogo};
        \end{tikzpicture}
      }{8pt}%
    }%
  }%
}

\newcommand{\Tr}{\text{Tr}}
\makeatletter

\begin{document}
\title{Scrutinizing the Mori memory function for diffusion in periodic quantum systems}
\author{Scott D. Linz~\orcid{0009-0005-7777-7955}}
\email{sclinz@uni-osnabrueck.de}
\affiliation{Department of Mathematics/Computer Science/Physics, University of Osnabr\"uck, D-49076 
Osnabr\"uck, Germany}

\author{Jiaozi Wang~\orcid{0000-0001-6308-1950}}
\affiliation{Department of Mathematics/Computer Science/Physics, University of Osnabr\"uck, D-49076 
Osnabr\"uck, Germany}

\author{Robin Steinigeweg~\orcid{0000-0003-0608-0884}}
\affiliation{Department of Mathematics/Computer Science/Physics, University of Osnabr\"uck, D-49076 
Osnabr\"uck, Germany}
\author{Jochen Gemmer~\orcid{0000-0002-4264-8548}}
\affiliation{Department of Mathematics/Computer Science/Physics, University of Osnabr\"uck, D-49076 
Osnabr\"uck, Germany}
\date{\today}

\begin{abstract}
Diffusion is an ubiquitous phenomenon . It is a widespread belief that as long as the area under a current autocorrelation function converges in time, the corresponding spatiotemporal density dynamics should be diffusive. This may be viewed as a result of the combination of linear response theory with the Einstein relation. However, attempts to derive this statement from first principles are  notoriously challenging. We first present a counterexample by constructing  a correlation functions of some density wave, such that the area under the corresponding current autocorrelation function converges, but the dynamics do not obey a diffusion equation. Then we will introduce a method based on the recursion method and the Mori memory formalism, that may help to actually identify diffusion. For a decisive answer,  one would have to know  infinitely many so called Lanczos coefficients, which is unattainable in most cases. However, in the examples examined in this paper, we find that the practically  computable number of  Lanczos coefficients suffices for a strong guess.
\end{abstract}

\maketitle

\section{Introduction}

Physical currents can arise from two distinct scenarios. Firstly, they can be triggered by some external force. In this case, linear response theory \cite{kubo2012statistical} states that the current $\hat{J}$ will be proportional to the external force $\hat{F}$ on the transported quantity,
\begin{equation}\label{1}
    \hat{J} = \chi \hat{F},
\end{equation}
where $\chi$ is the corresponding susceptibility. The other case, which will be the one of interest in this paper, arises from spatially inhomogeneous distributions of some density wave. In this case it is phenomenologically known that the following diffusion flux holds in one dimension,
\begin{equation}\label{diffusioneq}
   \hat{J}(t,x) = - D \frac{\partial}{\partial x} \hat{\varrho}(t,x),
\end{equation}
where $D$ is the diffusion constant and $\hat{\varrho}$ is the density of the transported quantity. This is known as Fick's first law.  The statement that (\ref{diffusioneq}) holds if (\ref{1}) holds with  a linear relation between $\chi$ and $D$ is well known  as the Einstein relation.  Applying the continuity equation to (\ref{diffusioneq}) leads to the diffusion equation
\begin{equation} \label{heat}
\frac{\partial}{\partial t} \hat{\varrho}(t,x) = D \frac{\partial^2 }{\partial x^2} \hat{\varrho}(t,x).    
\end{equation}
The direct derivation of (\ref{diffusioneq}) or (\ref{heat}) is more subtle than the derivation of  (\ref{1}). This is especially crucial for Fourier's law of heat conduction since in this case there is no possibility for any external force to cause a heat current. A set of approaches has been attempted to examine possible derivations (\ref{diffusioneq}, \ref{heat}), which are summarized in a review article by Zwanzig \cite{zwanzig1965time}. The paper at hand  will thoroughly investigate these subtleties.\\
First, operators for a density wave and a corresponding current will be defined, and we will examine their autocorrelation functions and discuss what diffusion implies for the dynamics of such autocorrelation functions. It will be shown that the dynamical evolution for the autocorrelation function of density wave can be cast into a linear integral Volterra equation or integro-differential equation. The memory kernel within this equation is found to have  similarities to the current autocorrelation function, but there is a subtle difference for nonzero wave numbers. Then the main question of this work will be examined, namely whether or not a finite area under a current autocorrelation function implies diffusive behavior \cite{forster2018hydrodynamic}. It will be shown, that counterexamples to the aforementioned claim can be constructed on the basis of specially designed correlation functions. This implies that the Einstein relations would be violated if a real system were to follow these dynamics.\\
To this end, both Lanczos' algorithm \cite{viswanath1994recursion} and the Mori memory formalism are introduced. Both are equivalent ways of analyzing quantum many-body systems at infinite temperature \cite{bartsch2024estimation}. In both cases, a collection of real numbers called the Lanczos coefficients play a crucial role. The Mori memory formalism \cite{mori1965transport, herbrych2012spin}, where the dynamics of autocorrelation functions is cast into a nested set of Volterra equations, will also play a central role in examining the memory dynamics. This formulation allows us to examine the precise difference between the memory function mentioned above and the current autocorrelation function. The Lanczos coefficients allow us to formulate an additional necessary criterion for diffusion, namely a condition under which the memory dynamics are approximately Markovian. \\
This paper is organized as follows: Section \ref{Basics} focuses on giving a basic intuition as to why a convergent area under a current autocorrelation function suggests diffusive behavior, but also demonstrates the loophole in this line of reasoning. A counterexample to this claim will be given. Section \ref{sec3} will provide the main tools that will be used to address this problem and will conclude with a necessary criterion for diffusion. In Section \ref{sec4}, the Lanczos coefficients for different scenarios will be examined numerically on the level of the autocorrelation functions. This analysis will demonstrate the usefulness of the above necessary condition.  Physical systems, namely spin chains, will also be analyzed, demonstrating that these systems fulfill our necessary condition, which is in line with the consensus that the systems in question are diffusive in the regime of the parameters chosen.

\section{Preliminary considerations} \label{Basics}

\subsection{Density wave and current operators}

In our approach to clarify conditions for diffusive behavior we will consider density waves. Let us consider a local observable, $\hat{A}_n$, which is centered around the site $n$ of a periodic system of condensed matter type, like, e.g., spin chain. We will furthermore assume that the sum over such operators on a lattice is a constant of motion
\begin{equation}\label{conserved}
    \frac{d}{dt} \sum_n \hat{A}_n = 0.
\end{equation}
A normalized density wave is thereby given as
\begin{equation} \label{Wave}
    \hat{\varrho}_k \coloneq \frac{1}{\sqrt{Z}} \sum_{n=1}^N  \hat{A}_n \cos(kn).
\end{equation}
Here, $k$ is the wave number and $N$ refers to the total number of sites. $Z$ is a normalization constant that will be specified below. It should be noted that for a given wave number $k$ the total number of sites takes on values that are multiples of the associated wavelength $\lambda =\frac{2\pi}{k}$.\\
To calculate the time evolution of some operator we make use of Heisenberg's equation, where we introduce the Liouville superoperator as $\mathcal{L} \hat{A} \coloneq [\hat{H}, \hat{A}]$ that maps some operator $\hat{A}$ to another operator. Here the Hamiltonian $\hat{H}=\sum_n \hat{h}_n$ shall be compromised of local sub-Hamiltonians $\hat{h}_n$, that only have a 2-local overlap with the observable $\hat{A}_n$. The introduction of the Liouville superoperator allows us to formulate Heisenberg's equation of motion in a compact form
\begin{equation} \label{Heisenberg}
    \frac{d}{dt} \hat{\varrho}_k = i\mathcal{L} \hat{\varrho}_k.
\end{equation}
We now argue that the derivative with respect to time of the density wave operator can be linked to some current. By applying (\ref{Heisenberg}) to the density wave and restricting ourselves to the case, that our Hamiltonian and the observable in consideration only overlap on at most two sites, 
we can write
\begin{equation}\label{deriv}
    \frac{d}{dt} \hat{A}_n(t) =  i [\hat{h}_{n-1},\hat{A}_n] + i [\hat{h}_n, \hat{A}_n] + i [\hat{h}_n, \hat{A}_{n+1}].
\end{equation}
From Eq.\ (\ref{conserved}), we know that the sum over the local operators is constant of motion. Furthermore, we modify our expression by rearranging our sum so that the last term in (\ref{deriv}) is shifted to the next part of the sum. This implies shifting the dummy index by $-1$ for this particular term. Thus,
\begin{equation}
    0 = \sum_n \left\{ [\hat{h}_{n-1}, \hat{A}_n] + [ \hat{h}_n, \hat{A}_n] + [ \hat{h}_n, \hat{A}_{n-1}] \right\}.
\end{equation}
must hold for Eq.\ (\ref{conserved}) to hold. For this sum to vanish all contributions on the different sub-Hilbert spaces associated with product spaces of different sites must be $0$. Therefore, we can see that 
\begin{equation} \label{comm2}
[\hat{h}_n, \hat{A}_n] = 0    
\end{equation}
and $[\hat{h}_{n-1}, \hat{A}_n] + [\hat{h}_n, \hat{A}_{n-1}]=0$ since these are the only terms defined on a given sub-Hilbert space. This was here demonstrated for the interaction of neighboring sites, but holds similarly for any 2-local interaction. So a corresponding argument is also valid for next nearest-neighbor interaction etc.\ which will be important for some the numerical simulations later on. Thereby we can identify
\begin{equation}\label{commut}
    [\hat{h}_n, \hat{A}_{n-1}] = - [ \hat{h}_{n-1}, \hat{A}_n].
\end{equation}
With (\ref{comm2}) and (\ref{commut}) we write out the time derivative of the density wave (\ref{Heisenberg}) as
\begin{equation} \label{cosines}
    \frac{d}{dt}\hat{\varrho}_k = \frac{1}{\sqrt{Z}} \sum_n i [\hat{h}_n,\hat{A}_{n+1}]\left\{ \cos(nk)-\cos[(n-1)k]\right\},
\end{equation}
where the sum again was rearranged so that only one commutator appears per summand. One can now link the density wave to a current $\hat{J}_k'$, which will only be used temporarily, by the continuity equation
\begin{equation}\label{continuity}
   \frac{d}{dt} \hat{\varrho}_k = -\frac{\partial}{\partial n} \hat{J}'_k,
\end{equation}
where $\frac{\partial}{\partial n}$ refers to the spatial derivative, which in our set-up is the change between neighboring sites. In the limit of small wave numbers we can rewrite the trigonometric weights in (\ref{cosines}) according to
\begin{equation} \label{cosines2}
    \cos(nk) - \cos[(n-1)k] \approx \frac{d \cos(nk)}{dn} = -k \sin(nk).
\end{equation}
Since we assume that our system is translationally invariant, the derivatives with respect to the sites of the local operators vanish. 
Thus we can define the following current, so that the continuity equation is fulfilled,
\begin{equation} \label{current}
    \hat{J}_k' \coloneq \frac{i}{\sqrt{Z}} \sum_n [\hat{h}_n,\hat{A}_{n+1}] \cos(kn).
\end{equation}
We have now defined a density wave and an associated current. In the following we will turn to the relationship between their respective autocorrelation functions.

\subsection{Autocorrelation functions of the density wave and current}

In our line of reasoning we will focus on correlation functions of some observable $\hat{O}$ at infinite temperature, that will be defined as
\begin{equation}
O(t) \coloneq \Tr[ \hat{O}^\dagger(0) \hat{O}(t)] / \Tr[\mathbb{1}],
\end{equation}
where the missing hat indicates that the quantity under discussion is a correlation function and the unity matrix $\mathbb{1}$ shall have the dimension of the operator Hilbert space in question. Following this line of reasoning the normalization constant $Z$ for the density wave $\varrho_k$ from (\ref{Wave}) will be chosen so that
\begin{equation}
\varrho_k(0)=1.
\end{equation}
We can define a new current operator $\hat{J}_k$ that will be used throughout the rest of this paper whose phase is shifted by $-\pi/2$,
\begin{equation} \label{newcurrent}
    \hat{J}_k \coloneq i \sum_n [\hat{h}_n, \hat{A}_{n+1}] \sin(nk).
\end{equation}
This second current has the property that its correlation is identical to the one for $\hat{J}_k'$, because translation invariance is assumed for the systems in consideration,
\begin{equation}\label{CurrCorr}
    J_k(t) \coloneq \Tr[\hat{J}_k(t) \hat{J}_k(0)] = \Tr[\hat{J}_k'(t)\hat{J}_k'(0)]. 
\end{equation}
With this definition and following (\ref{Heisenberg}) the second derivative of the correlation function of the density wave with respect to time can be written out as
\begin{align} \label{ddot}
    \ddot{\varrho}_k(t) &= - \Tr \left[ \hat{\varrho}_k(t) \mathcal{L}^2 \hat{\varrho}_k(0) \right] \\ \label{ddot2}
    &= -\Tr \left[ i \mathcal{L} \hat{\varrho}_k(t) i \mathcal{L} \hat{\varrho}_k(0) \right].
\end{align}
(The extra minus sign arises from shuffling the imaginary Liouvillian $i \mathcal{L}$ to the other side, since it is skew-Hermitian.) Equations (\ref{cosines}), (\ref{cosines2}) and (\ref{newcurrent}) lead to the identification of (\ref{Heisenberg}) as
\begin{equation} \label{Fund}
    \frac{1}{k} i \mathcal{L} \hat{\varrho}_k = \hat{J}_k.
\end{equation}
The definition of the new current (\ref{newcurrent}) can now be inserted into (\ref{ddot}) and (\ref{ddot2}) to yield a relationship between the correlation function for the density wave and an associated current,
\begin{equation}\label{Intro1}
    \frac{1}{k^2} \ddot{\varrho}_k(t) = -J_k(t).
\end{equation}
This expression links the correlation function for the density wave to the current autocorrelation function. This will be of importance for the rest of this paper, since the relation between the dynamics of the autocorrelation function for a density wave $\varrho_k(t)$ and for the current $J_k(t)$ will be examined.\\
Calling a process diffusive is the same as saying that its respective density correlation function decays exponentially, which will be demonstrated in the following.\\
We consider an initial state with a density matrix $\hat{d}$ of the form
\begin{equation}
\hat{d} \propto e^{\varepsilon \hat{\varrho}_k} = 1 + \varepsilon \hat{\varrho}_k + \mathcal{O}(\varepsilon^2)
\end{equation}
which is natural in linear response theory for high temperatures. Thereby, one arrives at the result that correlation functions behave very similarly to expectation values,
\begin{equation} \label{expect}
    \langle \hat{\varrho}_k(t) \rangle \propto \Tr[\hat{\varrho}_k(t)] + \varepsilon \Tr[\hat{\varrho}_k(0) \hat{\varrho}_k(t)].
\end{equation}
The trace over the first term vanishes. Therefore autocorrelation functions show identical behavior to expectation values, i.e.
\begin{equation} \label{expect2}
    \langle \hat{\varrho}_k(t) \rangle = \varepsilon \Tr [\hat{\varrho}_k(t) \hat{\varrho}_k(0)] \propto \varrho_k(t).
\end{equation}
If the diffusion equation (\ref{heat}) is Fourier transformed in space, it reads,
\begin{equation} \label{DiffOpEq}
    \frac{\partial \hat{\varrho}_k(t)}{\partial t} = - D k^2 \hat{\varrho}_k(t),
\end{equation}
which means a harmonic density wave will have an (mono-)exponential decay in time as can be seen by the solution of this differential equation
\begin{equation}\label{TimeOp}
    \hat{\varrho}_k(t) = \hat{\varrho}_k(0) e^{-Dk^2 t}.
\end{equation}
Thus , (\ref{TimeOp}) is the solution to the diffusion equation (\ref{heat}). Given a diffusive process the temporal correlation function can thereby be written as,
\begin{equation}\label{diffusion}
    \varrho_k(t) = e^{- D k^2 t}.
\end{equation}
 Only if this holds for all small $k$ the corresponding process is diffusive. 

\subsection{Intuition and counterexample}

We will present the intuitive argument concerning the assumption that a finite area under a current autocorrelation function would imply  a diffusive process. Since a link between a density wave and a current has been established for both the operators and correlation functions, our investigation will turn to the Laplace transforms of both the density wave and the current with respect to time. As long as a Laplace transform for some function exists, it is always possible to cast the dynamics of this function into an integro-differential equation,
\begin{equation}\label{integrodiff}
    \dot{\varrho}_k(t) = - k^2 \int_0^t dt' M_k(t-t') \varrho_k(t'),
\end{equation}
where $M_k(\tau)$ is some as of yet unspecified memory kernel. Following this train of thought, there exists a bijective mapping between the Laplace transform of $\varrho_k(t)$ and the Laplace transform of the memory function. By setting $\varrho_k(0)=1$ this yields,
\begin{equation}
    \tilde{\varrho}_k(s) = \frac{1}{s+k^2 \tilde{M}_k(s)} \label{Laplace}
\end{equation}
Here, the tilde indicates that we are making use of the Laplace transform in time of the correlation function under discussion. The variable $s$ represents the frequency parameter. Laplace transforming (\ref{Intro1}) formally yields 
\begin{equation}\label{lapcurr}
     \frac{1}{k^2} \ddot{\tilde{\varrho}}_k(s) = -\tilde{J}_k(s).
\end{equation}
Using basic Laplace calculus  and setting $\dot{\varrho}_k(0) =0$, one arrives from combining (\ref{Laplace}) and  (\ref{lapcurr}) at
\begin{equation}
    \tilde{J}_k(s) = \frac{s \tilde{M}_k(s)}{s+k^2 \tilde{M}_k(s)}.
\end{equation}
Boldly setting $k=0$ one obtains
\begin{equation}
    \tilde{J}_0(s) = \tilde{M}_0(s),
\end{equation}
which would precisely mean that the current autocorrelation function is the memory kernel for the integro-differential equation (\ref{integrodiff}) for the wave's correlation function. A conclusion that could be drawn from this would be that
\begin{equation}
    \tilde{M}_{k \rightarrow 0} (s) \approx \tilde{M}_0(s) = \tilde{J}_0(s).
\end{equation}
which  entails $M_k(t)\approx J_0(t)$ in the limit of small $k$. Inserting this into (\ref{integrodiff}) and assuming  that $J_0(t)$  decays completely on a finite time scale yields  $\varrho_k(t) \approx  e^{-D k^2 t}$, again in the limit of small $k$. Therefore, a finite area under current autocorrelation with respect to time at $k=0$ would produce a necessary and sufficient criterion for diffusive behavior.\\

However, this approach includes an unwarranted assumption, namely that the transition from small $k$ to $k=0$ is trivial. In the case of a vanishing wave number, the dynamics of the density wave becomes trivial, since then by (\ref{integrodiff}) the density "wave" is a conserved quantity. But for any $k \neq 0$ we will have nontrivial dynamics in Eq.\ (\ref{integrodiff}). Furthermore, the limit of very small $k$ is of particular interest since it coincides with the regime of long wave lengths, where we would expect the hydrodynamic diffusion equation to hold.\\
To illustrate why this can cause a problem, a counterexample will be given, where the area under the current autocorrelation function is finite; however, the process is not diffusive. First we will consider a wave whose correlation decays according to
\begin{equation}
    \varrho_k(t) \coloneq \frac{1}{1+k^2|t|}. \label{Corr}
\end{equation}
One can compute the current autocorrelation function according to Eq.\ (\ref{Intro1}),
\begin{equation}\label{Finite}
    J_k(t) = 2 \left( \delta(t) - \frac{k^4}{(1+k^2|t|)^3} \right).
\end{equation}
If we now set our wave number to $k=0$, we simply obtain a delta function, which implies decay over a finite time scale. As seen in (\ref{Corr}), the correlation function of the wave does not decay exponentially. This in turn means that the behavior is not diffusive. It implies that purely on the basis of correlation functions, we are able to construct a counterexample to the claim that a finite area under a current autocorrelation function implies a diffusive decay for the correlation function of the associated wave. This counterexample is by no means unique. Many more may be constructed along the same lines. 

\section{Lanczos algorithm and the Mori memory formalism applied to diffusion}\label{sec3}

\subsection{Lanczos algorithm}\label{Lanczos2}

Here, the Lanczos algorithm will be introduced \cite{viswanath1994recursion}, since the Lanczos coefficients arising in this scheme will be important for a necessary criterion for whether or not a process is diffusive. Knowing all of the Lanczos coefficients generated by a physical system is equivalent to knowing the correlation function of the system. In most cases only a finite number of these can be calculated. These $n$ Lanczos are equivalent to the first $2n$ moments of the correlation function.
Before the Lanczos algorithm is formulated, we will introduce new notation by viewing operators as vectors in a Hilbert space. In elementary quantum mechanics a state vector is mapped to another state vector by an operator. Analogously one can view operators as vectors within an operator Hilbert space. These will then be denoted by $| \hat{O} )$ with its dual element $( \hat{O}|$. Between two such vectors one can define an inner product, which is also known as the Frobenius scalar product by
\begin{equation}
( \hat{O}_1 | \hat{O}_2 ) \coloneq \Tr[ \hat{O}_1^\dagger \hat{O_2}]/\Tr[\mathbb{1}].
\end{equation}
This represents the Kubo scalar product at infinite temperature \cite{kubo2012statistical}. Moreover a norm is induced by this inner product that is given by $|| \hat{O} || = \sqrt{( \hat{O} | \hat{O} )}$.
For a given observable $\hat{O}$ in the Heisenberg picture and a Hamiltonian $\hat{H}$, that governs its time evolution, the correlation function of the observable can be defined as
\begin{equation}
O(t) = ( \hat{O}(t) | \hat{O}(0) ).
\end{equation}
The time evolution of the operator in consideration is obtained by $\hat{U}^\dagger (t) \hat{O} \hat{U}(t)$, where $\hat{U}(t) = \exp(-i H t)$. To describe the dynamical evolution of a system it is furthermore useful to make use of the previously introduced Liouville superoperator. Given the Heisenberg equation of which (\ref{Heisenberg}) is a special case, one can write
\begin{equation}
\dot{\hat{O}} = i \mathcal{L}\hat{O}.
\end{equation}
Since the Liouvillian describes the time evolution of a given operator the correlation function can then also be written as
\begin{equation} \label{Correl}
O(t) = ( \hat{O} | e^{i \mathcal{L} t} | \hat{O} ). 
\end{equation}
The Lanczos algorithm, as discussed in this section, is an algorithm to generate an orthonormal basis in which the Liouville superoperator takes a tridiagonal form. Summarized briefly, one generates the Krylov basis, which is the set of operators where the Liouvillian is applied to the seed operator up to $n$ times successively. By (\ref{Heisenberg}) this is the same as computing the $n$-th derivative with respect to time of the original seed operator. This Krylov basis of the Liouvillian will then be orthonormalized a la Gram-Schmidt. The set of normalization constants, that arise within this scheme, are called the Lanczos coefficients and will be denoted by $b_n$. To formulate this procedure in mathematical terms we start with some normalized seed operator $\hat{O} = | 0^* )$ and apply the Liouvillian to it, giving us $| 1 ) = \mathcal{L} | 0^* )$. Its norm $b_1 = \sqrt{ ( 1 | 1 )} $ is the first Lanczos coefficient. Then one normalizes the resulting operator, yielding $| 1^* )$, where the star indicates normalization. The successive steps in this algorithm can then be formulated recursively. To generate the $n$-th Lanczos coefficient $b_n$ and basis operator $| n )$, one must follow through with the following steps,
\begin{align}
| n ) &= \mathcal{L} | [n-1]^* ) - b_{n-1} | [n-2]^* ) \label{Lanczos} \\
b_n^2 &= ( n | n ) \\
| n^* ) &= | n ) / b_n. \label{Normal}
\end{align}
If one inserts (\ref{Normal}) into the l.h.s. of (\ref{Lanczos}), one arrives at
\begin{equation}
    b_n | n^* ) = \mathcal{L} |[n-1]^* ) - b_{n-1} | [n-2]^* ).
\end{equation}
By increasing the numbering by one and rearranging, we arrive at
\begin{equation}
    \mathcal{L} |n^* ) = b_{n+1} |[n+1]^* ) + b_n | [n-1]^* )
\end{equation}
Multiplying by an element of the dual basis one then arrives at the matrix representation of the Liouvillian. With the constructed basis the Liouvillian then takes a tridiagonal form,
\begin{equation}
\mathcal{L}_{nm} = ( n^* | \mathcal{L} | m^* ) = \begin{pmatrix}
0 & b_1 & 0 & 0 & \hdots \\
b_1 & 0 & b_2 & 0 & \hdots \\
0 & b_2 & 0 & b_3 &  \\
0 & 0 & b_3 & 0 & \ddots \\
\vdots & \vdots & & \ddots & \ddots
\end{pmatrix}.
\end{equation}
A special feature of the Lanczos algorithm in the Liouvillian representation is that all diagonal elements vanish ($\mathcal{L}_{nn} = 0$). This feature results from the nature of the Frobenius inner product and can be simply verified by applying the fact that cyclic permutations within a trace yield the same trace.\\
\begin{figure}
  \centering
	\includegraphics[scale=0.4]{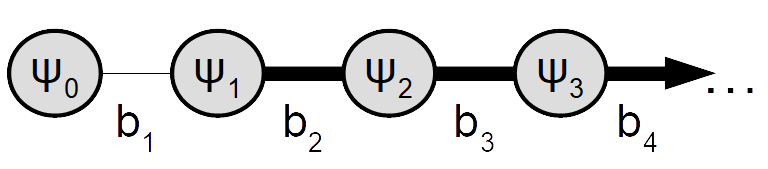}
	\caption{Here the semi-infinite chain is illustrated. The $\psi_n$'s are the coefficients of  some operator $\hat{O}$ when expanded w.r.t. the "local"  Lanczos vectors  $|n^*)$.  The Lanczos coefficients are the hopping amplitudes. The chain starts at "site" $|0^*)$ and it extends infinitely to the right for an infinite system. The magnitude of the hopping amplitudes (Lanczos coefficients) between sites is signified by the thickness of the connecting line. Here the first Lanczos coefficient is small while the others are larger, which will show up in Section \ref{Difference}.
    }
    \label{pikto}
\end{figure}
The Lanczos algorithm, being highly abstract, begs for a more intuitive illustration. One can simply interpret the Lanczos coefficients as hopping amplitudes in a semi-infinite chain, as illustrated in Fig.\ \ref{pikto} \cite{parker2019universal}.

\subsection{From Lanczos to Mori memory formalism}\label{Moriform}

To cast the dynamics in a form similar to (\ref{integrodiff}), we transfer from the Lanczos algorithm to the Mori memory formalism \cite{wang2024diffusion}. These two ways of interpreting the dynamics of a correlation function lead to equivalent results \cite{bartsch2024estimation}. The Lanczos algorithm leads to a infinite set of nested first-order differential equations, while the Mori memory formalism leads to an infinite set of Volterra equations, where each following equation describes the dynamics of the memory kernel of the previous equation. The resulting nested equations will be given in (\ref{volt}) and (\ref{Memkern}).\\
To start with we introduce  sub-Liouvillians, by erasing the first $n$ rows and columns,
\begin{equation} \label{SubLiou}
    \mathcal{L}_n = \sum_{m=n+1} b_m \left( |[m+1]^* ) ( m^* | + | m^* ) ( [m+1]^* | \right).
\end{equation}
 where the full Liouvillian is just $\mathcal{L}_0 \coloneq \mathcal{L}$. By construction for any $n$, the sub-Liouvillian $\mathcal{L}_n$ will be hermitian. Therefore let us write out its eigenstate equation by denoting the eigenvectors as $| q_n^* )$,
\begin{equation}
    \mathcal{L}_n | q_n^* ) = E_{q_n} | q_n^* )
\end{equation}
with the $|q_n^*)$ being orthonormal. Focusing on the first sub-Liouvillian $\mathcal{L}_1$ and by making use of a basis including $| 0^* )$, $| 1^* )$ and the eigenvectors $| q_n^* )$, this entire Liouvillian can be written out as
\begin{align}
    \mathcal{L} &= \sum_{q_1} b_1 \left\{ | q_1^* ) ( q_1^* | 1^* ) ( 0^* | + | 0^* ) ( 1^* | q_1^* ) ( q_1^* |  \right\} \\
    &+ E_{q_1} | q_1^* ) ( q_1^* |
\end{align}
So far our analysis was strictly applied within an analogue of the the Schr\"odinger picture which suggests the notation $\psi_n(t):= (n^*| e^{i \mathcal{L}_0 t} | 0^* ) $ and $|\psi (t) ):=     \sum_n    \psi_n(t) |n^*)$, cf. Fig. \ref{pikto}. Now let us switch to the interaction picture by taking $\mathcal{L}_1 $ as the Hamiltonian of the unperturbed system, whereby the perturbation then reads
\begin{align}
\begin{split}
    \mathcal{V}_I(t) = \sum_{q_1} b_1 \{ | q_1^* ) e^{-i E_{q_1} t} ( q_1^* | 1^* ) ( 0^* | \\
    + |0^* ) ( 1^* | q_1^* ) e^{i E_{q_1}t} ( q_1^* | \}
\end{split}
\end{align}
The time evolution of the perturbation within the interaction picture is now governed by
\begin{equation}
i \frac{\partial}{\partial t}  | \psi_I (t) ) = \mathcal{V}_I(t) | \psi_I(t) ).
\end{equation}
With our aforementioned basis we can cast the dynamics into a new form,
\begin{align}
    \dot{r}_0 = -i b_1  \sum_{q_1} ( 1^* | q_1^* ) e^{i E_{q_1} t} r_{q_1} \label{Integrate2} \\
    \dot{r}_{q_1} = -i b_1 ( q_1^* | 1^* ) e^{-i E_{q_1} t} r_0, \label{Integrate}
\end{align}
with the shorthand notation
\begin{align}
    r_0 = ( 0^* | \psi_I(t) ), \\
    r_{q_1} = ( q_1^* | \psi_I (t) ).
\end{align}
In attempting to decouple these equation, it is a useful step to formally integrate (\ref{Integrate}), yielding
\begin{equation}
    r_{q_1}(t) = r_{q_1}(0) - i b_1 ( q_1^* | 1^* ) \int_0^t dt' r_0 (t') e^{-i E_{q_1} t}.
\end{equation}
For our purposes it is sufficient to specialize to cases with  $r_{q_1}(0) = 0$. This can now be inserted into (\ref{Integrate2}), which then reads
\begin{equation}
    \dot{r}_0 (t) = - b_1^2 \int_0^t dt' \sum_{q_1} |( q_1^* | 1^* ) |^2 e^{i E_{q_1}(t-t')} r_0(t').
\end{equation}
This is a standard Volterra equation, where one can simplify the notation by introducing a memory kernel,
\begin{equation} \label{Kernel}
    C_1(\tau) := \sum_{q_1} | ( q_1^* | 1^* ) |^2 e^{i E_{q_1} \tau} = ( 1^* | e^{i \mathcal{L}_1 \tau} | 1^* ).
\end{equation}
Thus, we write
\begin{equation}\label{kern}
    \dot{r}_0 (t) = - b_1^2 \int_0^t dt' C_1(t-t') r_0(t').
\end{equation}

Here it should be pointed out  that at $|0^*)$ the  coefficients of the wave function are the same in the Schr\"odinger and the interaction picture, i.e.,  $r_0(t) = \psi_0(t)$. To arrive at a concise notation, we give this quantity yet another name, $\psi_0(t):= C_0(t) $.  This turns  (\ref{kern}) into 
\begin{equation} \label{Original}
     \dot{C}_0(t) = - b_1^2 \int_0^t dt' C_1(t-t') C_0(t'). 
\end{equation}
Comparing  the definitions of  $C_0$ and $C_1$ we realize that that these are similar and both fit into the more general scheme 
\begin{equation}\label{Memkern}
    C_n(\tau) = ( n^* | e^{i \mathcal{L}_n \tau} | n^* ).
\end{equation}
We further realize that the whole consideration in this Sect. could be repeated to arrive at an integro-differential equation for $C_1(t)$ (instead of $C_0(t)$) which would be the same as (\ref{Original}) only with the index shifts $0\rightarrow 1, 1 \rightarrow 2$. This may be iterated to produce a hierarchy of nested Volterra equations 
\begin{equation}\label{volt}
    \dot{C}_n(t) = -b_{n+1}^2 \int_0^t dt' C_{n+1}(t-t') C_n(t').
\end{equation}
This is in principle an autonomous set of equations from which $C_0(t)$ may be determined, given that all $b_n$ are konwn.

Eq. (\ref{volt}) features a structure very similar to (\ref{integrodiff}) which is one main result of this paper. This similarity will be scrutinized in even more detail in the next Section. \\
The Laplace transform of any integro-differential equation of the form Eq.\ (\ref{integrodiff}) is always a fraction of the form of  (\ref{Laplace}).  Similarly Laplace transforming  (\ref{volt}) entails a recursive scheme for the $\tilde{C}_n(s)$ which has a solution for $\tilde{C}_0(s)$ in  form of a continued fraction \cite{haydock1980recursive},
\begin{equation}\label{Area}
    \tilde{C}_0(s) = \frac{1}{s + \frac{b_1^2}{s + \frac{b_2^2 }{ \ddots {s+ \frac{\ddots b_{n-1}^2}{\tilde{C}_n(s)}}}}},
\end{equation}
(The properties of terminating this continued fraction have been an area of research \cite{viswanath1990recursion, bartsch2024estimation}. ) A useful property of Laplace transforms is that by setting $s=0$, we obtain the area under the respective correlation function with respect to time. 

\subsection{Memory function versus current autocorrelation function}\label{Difference}

In this section, the goal is to show how the current autocorrelation function $J_k(t)$ and the memory kernel $M_k(t)$ for the density wave differ.\\
The basis operators corresponding to our transport scenario within the framework of the Lanczos algorithm will be denoted by $|n,k)$ to signify the wave number dependence. Here, our normalized density wave given by (\ref{Wave}) is chosen as
\begin{equation}\label{X}
|[0,k]^* ) \coloneq \hat{\varrho_k}.
\end{equation}
The correlation function for the density wave, which can by be identified as $C_0(t)=\varrho_k(t)$ along the lines of (\ref{Original}), is
\begin{equation}
\varrho_k(t) = ( [0,k]^* | e^{i \mathcal{L} t} | [0,k]^* ).
\end{equation}
Following the framework introduced by Parker et al. \cite{parker2019universal} and visualized in Fig. \ref{pikto}, we can interpret the operator dynamics as those of a particle hopping on a 1D lattice that has an edge on the left side. Here $( [0,k]^* | e^{i \mathcal{L} t} | [0,k]^* )$ is the amplitude for the particle being at the first site at time $t$, if the particle was initialized on the first site. The initial conditions for this case are $\psi_0(0)=1$ and $\psi_n(0)=0$ for any natural number $n \geq 1$.\\
By taking Eq. (\ref{Fund}) into account, applying the Liouvillian and then normalizing for the first recursive step yields 
\begin{equation} \label{1k}
|[1,k]^* ) \coloneq -i \hat{J}_k  / ||\hat{J}_k(0)||.
\end{equation} 
For the current autocorrelation function we arrive at,
\begin{equation}\label{CurrentCorr}
J_k(t) = ||\hat{J}_k(0)||^2 ( [1,k]^*| e^{i \mathcal{L} t} | [1,k]^* ),
\end{equation}
where $( [1,k]^*| e^{i \mathcal{L} t} | [1,k]^* )$ describes the amplitude of the particle at the second site  at time $t$ if the particle is initially at the second site. For our hopping chain, this would imply the initial conditions $\psi_0(0)=0$, $\psi_1(0)=1$ and $\psi_n(0)=0$, where $n$ is any natural number $n\geq 2$.
Note that according to  (\ref{CurrCorr}) the dynamics of $J_k(t)$ is generated by the entire Liouvillian.\\
\begin{table}
\begin{tabular}{|c|c|c|c|c|c|c|} \hline
     & $\mathcal{L}_n$ & $\psi_0$ & $\psi_1$ & $\psi_2$ & \dots & $\psi_n$ \\ \hline
     $\varrho_k(0)$ &  & $1$ & $0$ & $0$ & \dots & $0$ \\ \hline
     $\varrho_k(t)$ & $\mathcal{L}_0$ & $\psi_0(t)$ & $\psi_1(t)$ & $\psi_2(t)$ & \dots & $\psi_n(t)$ \\ \hline
     $J_k(0)$ & & $0$ & $1$ & $0$ & \dots & 0 \\ \hline
     $J_k(t)$ & $\mathcal{L}_0$ & $\psi_0(t)$ & $\psi_1(t)$ & $\psi_2(t)$ & \dots & $\psi_n(t)$ \\ \hline
     $M_k(0)$ & & $0$ & $1$ & $0$ & \dots & $0$ \\ \hline
     $M_k(t)$ & $\mathcal{L}_1$ & 0 & $\psi_1(t)$ & $\psi_2(t)$ & \dots & $\psi_n(t)$ \\ \hline
\end{tabular}
\caption{Here the sub-Liouvillians according to which the time evolution is generated are listed in the first column. The initial conditions and time-evolution of the hopping chains illustrated in Fig. \ref{pikto} are displayed for ease of comparison in the following columns. The row represent the dynamical quantities of this interest throughout this paper. Furthermore, the rupture of the first site for the memory kernel dynamics is illustrated in the last row.}
 \label{Tb}
\end{table} 
It is instructive to calculate the first Lanczos coefficient. Here (\ref{Fund}), (\ref{X}) and (\ref{1k}) are used and this results in
\begin{align} \label{Coeff2}
b_1 (k) &= ( [1,k]^* | \mathcal{L} | [0,k]^* ) = i ( \hat{J}_k| -ki | \hat{J}_k ) / ||\hat{J}_k(0)||\\
&= k ||\hat{J}_k(0)||.\label{Coeff}
\end{align}
For later reference note that $b_1(k) \propto k$, i.e. it vanishes in the limit of small $k$.
Eqs.\ (\ref{Coeff2}) and (\ref{Coeff}) allow us to see that the first Volterra equation (\ref{Original}) of Mori memory formalism yields an expression of the form (\ref{integrodiff}). The memory function is just the memory function from (\ref{Memkern}) if one sets $n=1$ and normalizes according to
\begin{equation} \label{NormalKern}
C_1(t) = M_k(t) / ||\hat{J}_k(0)||^2.    
\end{equation}
All of this lets us rewrite the memory kernel from Eq.\ (\ref{integrodiff}) as
\begin{align}
M_k(t) &= ||\hat{J}_k(0)||^2( [1,k]^* | e^{i \mathcal{L}_1 t} | [1,k]^* ). \\
&= \frac{b_1^2}{k^2} ( [1,k]^* | e^{i \mathcal{L}_1 t} | [1,k]^* ). \label{b1}
\end{align}
An explicit representation of $\mathcal{L}_1$ that governs the time evolution of $M_k(t)$ can be given by
\begin{equation} \mathcal{L}_1 = 
    \begin{pmatrix}
        0 & 0 & 0 & 0 & \hdots \\
        0 & 0 & b_2(k) & 0 & \hdots \\
        0 & b_2(k) & 0 & b_3(k) &  \\
        0 & 0 & b_3(k) & 0 & \ddots \\
\vdots & \vdots & & \ddots & \ddots
    \end{pmatrix}.
\end{equation}
Returning to the intuition of the semi-infinite hopping chain, with $\mathcal{L}_1$ governing the time evolution of $M_k(t)$ the connection to the first site for $M_k(t)$ is broken. Thereby, $( [1,k]^* | e^{i \mathcal{L}_1 t} | [1,k]^* )$ is the amplitude at the second site at time $t$, if we initialize at the second site, an let it evolve with no connection to the first. For our hopping chain, this would imply $\psi_0(t)=0$ for all $t$, $\psi_1(0)=1$ and $\psi_n(0)=0$, where $n$ can be any natural number $n \geq 2$. (All statements made in this subsection regarding the hopping chain are presented in Table \ref{Tb} for ease of comparison.) Precisely the above  rupture of the connection to the first site is the difference between $M_k(t)$ and $J_k(t)$. If we set $b_1(k)=0$, we recover $J_0(t) = M_0(t)$. Now we could expect this to hold for the limit $M_{k \rightarrow 0} \approx M_0(t)$. However, we have already demonstrated that this cannot be generally correct, since we have already seen that counterexamples can be constructed. Whether or not particular physical systems exhibit the features of the counterexample, needs to be examined more closely, which motivates the next subsection.

\subsection{Necessary criterion for diffusion} \label{Conclusive}

In the previous Sections we have seen that a finite area under a current correlation function suggests diffusive behavior. But under closer inspection, one can see that the transition from a small wave number $k$ to the case $k=0$ is non-trivial. Therefore, merely examining the area under a current autocorrelation function is insufficient to establish if a system is diffusive. To make a distinction between a truly diffusive case or pathological behavior, additional criteria are needed. This motivates why we will give a necessary condition for diffusion, which can help to judge whether an example that has a finite area under the current autocorrelation function will truly be diffusive or not. However, it shall be noted that this is another  necessary but not sufficient condition for diffusive behavior.\\
If our dynamics yields diffusive behavior, then the memory function must induce Markovianity in Eq.\ (\ref{integrodiff}). This requires  a memory function that decays on a finite time scale. Thereby, Eq.\ (\ref{integrodiff}) would yield the differential equation of an exponential decay scaling in $k^2$. Decaying on a finite time scale would also mean that the area under the memory function is finite. We follow the framework laid out in Section \ref{Moriform}, where in Eq.\ (\ref{Area}) the Laplace transform of $C_0(t)$ is given. The Laplace transform of a function at the frequency parameter $s=0$ is simply the area under the function from $t=0$ to $\infty$. We can simply take (\ref{Laplace}) and (\ref{volt}) into consideration to see that (\ref{Area}) results for the Laplace transform of the memory function if we start with the second step in the recursion process, meaning all indices are increased by $1$. If we now set $s=0$, the area under the normalized memory function (\ref{NormalKern}) is simplified to an infinite product involving only Lanczos coefficients,
\begin{equation} \label{product}
    C_1(0) = b_2 \prod_{n=2}^\infty \left( \frac{b_{n}}{b_{n+1}} \right)^{(-1)^n}.
\end{equation}
Now, one only has to figure out whether or not this infinite product given by Eq.\ (\ref{product}) converges to a finite value as $n \rightarrow \infty$. If this product converges, we can be sure that the area under the memory function is finite, which leads to approximately Markovian, i.e. exponential,  dynamics. This is precisely our necessary condition for diffusion. The main issue with this test is that one can usually only calculate a finite number of Lanczos coefficients from an observable and a Hamiltonian for an arbitrary system. This number usually ranges from somewhere between $10$ and $40$ depending on the complexity of the operator and Hamiltonian in question.

\section{Numerical calculation for concrete examples} \label{sec4}

\begin{figure}[t]
\includegraphics[width=1\linewidth]{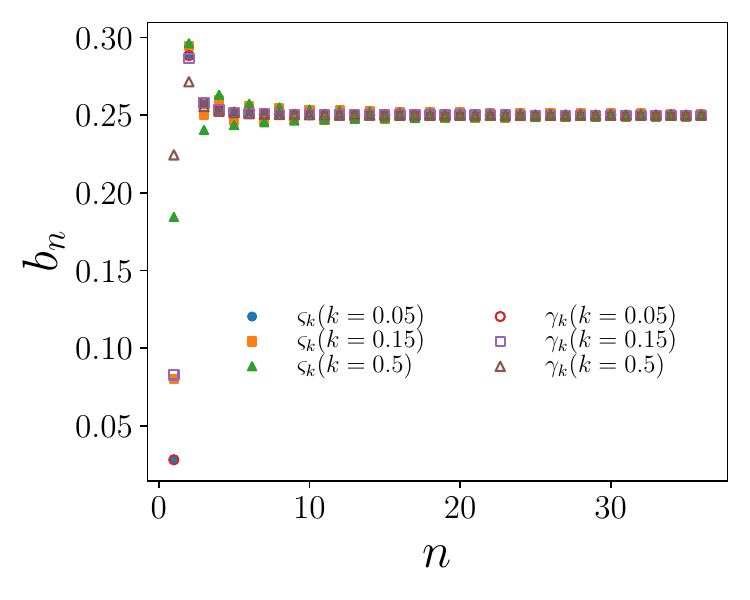} 
     \fontsize{15}{10}
    \caption{The Lanczos coefficients $b_n$ for the counterexample $\varsigma_k(t)$ as well as the diffusive exponential decay $\gamma_k(t)$ are plotted against $n$ up to $n=35$. In both cases different wave numbers were examined. Their values are $k_1 = 0.05$, $k_2 = 0.15$ and $k_3 = 0.5$}
    \label{fig2}
\end{figure}

\subsection{On the basis of correlation functions}

We have suggested in Section \ref{sec3} that whenever the product defined by (\ref{product}) is divergent, and thereby the area under the correlation function is not finite, a system cannot exhibit diffusive behavior. Since we already constructed a counterexample in Section \ref{Basics}, this divergence will now be investigated numerically. The method by which Lanczos coefficients are calculated from a correlation function is reviewed in the Appendix.
Here we will compare the Lanczos coefficients arising from our counterexample given by (\ref{Corr}). The counterexample will now be referred to as $\varsigma_k(t)$, whereas the exponentially decaying correlation function will be called
\begin{equation}
    \gamma_k(t) \coloneq \exp(-k^2|t|). \label{Exp}
\end{equation}
To make the identification simpler, the Lanczos coefficients and partial products will be marked either $\varsigma_k$ or $\gamma_k$ in the legends of our figures, depending on whether or not they arise from the counterexample or the exponential decay. An exponentially decaying correlation function of the form Eq.\ (\ref{Exp}) is equivalent to claiming that a system follows the diffusion equation, as stated in Section \ref{Basics}. In both cases, the first Lanczos coefficient, which is essentially the derivative with respect to time at $t=0$, is ill defined. Both functions are not differentiable at $t=0$. This, however, can be solved by applying convolution, with a sinc function
\begin{equation}
    F(t) * \text{sinc}(at) = \sqrt{2 \pi}\mathcal{F}^{-1}\left[F(\omega) \text{Rect}\left(\frac{\omega}{a}\right)\right],
\end{equation}
where $\mathcal{F}^{-1}$ is the inverse Fourier transform and $F(t)$ is some function with a Fourier transform $F(\omega)$. This means cutting off the edges in frequency space by multiplying by the rectangular function $\text{Rect}(\frac{\omega}{a})$ with width $a$. In the numerical simulations presented in this paper, we chose $a=1$.\\
\begin{figure}[t]
\includegraphics[width=1\linewidth]{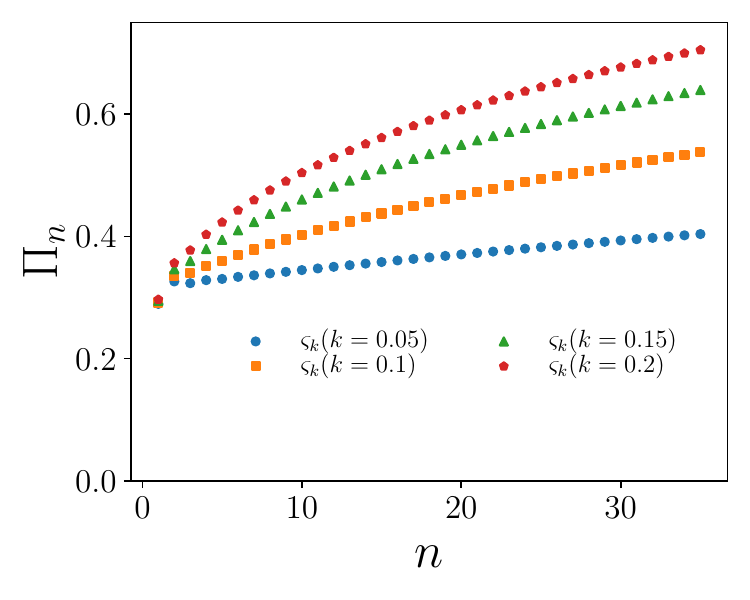} 
    \caption{The partial products $\Pi_n$ are plotted against $n$ for $\varsigma_k(t)$ for different wave numbers. Here we have plotted them for $k_1=0.05$, $k_2=0.1$, $k_3=0.15$ and $k_4=0.2$.}
    \label{fig3}
\end{figure}
\begin{figure}[t]
\includegraphics[width=1\linewidth]{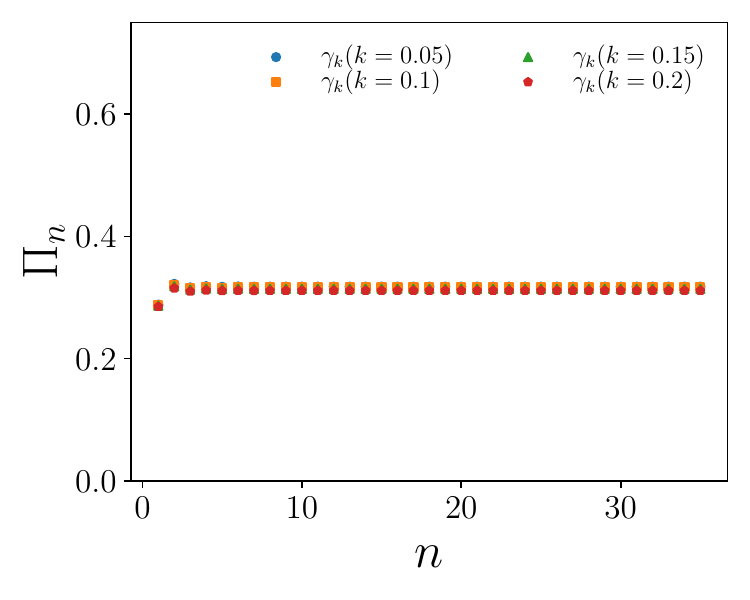} 
    \caption{The partial products $\Pi_n$ are plotted against $n$, but this time for $\gamma_k(t)$. For this case we have used the wave numbers $k_1=0.05$, $k_2=0.1$, $k_3=0.15$ and $k_4=0.2$.}
    \label{fig35}
\end{figure}
As can be seen in Fig.\ \ref{fig2} for both cases, we can observe that the first coefficient is small for small wave numbers, and then there is a leap. For an exponential decay, the Lanczos coefficients quickly approach a constant value. This tendency is also present for the counterexample. However, here there is also a slight even-odd effect present, where the even Lanczos coefficients are above the constant and the odd coefficients fall below it. This effect becomes more pronounced as the wave number increases.\\
The convergence or lack thereof of the infinite product given Eq.\ (\ref{product}) is central to our numerical analysis. Since we do not know all the Lanczos coefficients for our correlation function, we will investigate the behavior of the sequence of partial products
\begin{equation}
    \Pi_n \coloneq b_2 \prod_{i=2}^n \left( \frac{b_i}{b_{i+1}} \right)^{(-1)^i},
\end{equation}
where the infinite product in Eq.\ (\ref{product}) is cut off at $n$ and $\Pi_1 = b_2$, to examine whether a convergence becomes apparent. We can plot the partial products given by the formula (\ref{pin}) against $n$. These are plotted in Figs.\ \ref{fig3} and \ref{fig35} both for $\varsigma_k(t)$ and $\gamma_k(t)$ for different wave lengths. As we can see for the exponential decay $\gamma_k(t)$ in Fig. \ref{fig35} the partial products soon approach a value that is nearly constant. But for all the wave numbers $k$ examined here, the partial product for $\varsigma_k(t)$ continues to grow, as illustrated in Fig. \ref{fig3}. This growth is more pronounced for larger wave numbers. This is in line with our expectation that the partial product for $\varsigma_k(t)$ does not converge and we do not see diffusive behavior although the area under the current correlation function is finite as discussed in Eq.\ (\ref{Finite}).\\

\subsection{On the basis of physical systems}

Correlation functions generally arise from the dynamics of a given system, evolving according to some Hamiltonian. We have seen that one can construct a counterexample to the claim examined here on the level of correlation functions. Therefore, we will turn our attention to the behavior of real physical systems. These systems are in the diffusive regime according to the consensus in condensed matter physics. Therefore, they should not show pathological behavior and therefore the area under the correlation function should not diverge, as measured by the limit of Eq.\ (\ref{product}). Here, the Lanczos coefficients will be generated according to (\ref{Lanczos}) for both a tilted-field Ising model \cite{rakovszky2022dissipation, artiaco2024efficient, thomas2023comparing} and the Heisenberg XXZ model, where in both cases we only consider the parameters where diffusion is expected.\\
The Hamiltonian of the titled-field Ising model reads $H = \sum_{n=1}^N h_n$, where
\begin{equation}
h_n=4Js_{n}^{z}s_{n+1}^{z}+B_{x}s_{n}^{x}+B_{z}s_{n}^{z}\ .
\end{equation}
$s_n^i$ ($i = x,y,z$) are the components of a spin-$1/2$ operator at lattice site $n$, $N$ is the total number of lattice. We consider energy transport, ie.,  for the seed operator we choose an energy density wave like Eq.\ (\ref{Wave}), where $\hat{A}_n = h_n$
The parameters are chosen to be $J = 1.0,\ B_x = 2.0, \ B_z = 1.0$, which is within the diffusive regime. The Lanczos coefficients arising from this set-up have also been studied in \cite{noh2021operator} \\
The Hamiltonian $H=\sum_{n=1}^L h_n$ of the Heisenberg XXZ model can be written as a sum over of local Hamiltonians
\begin{equation}
h_n = s_{n}^{x} s_{n+1}^{x} + s_{n}^{y} s_{n+1}^{y} + \Delta
s_{n}^{z}s_{n+1}^{z} + \Delta^\prime s_{n}^{z}s_{n+2}^{z} \,
,
\end{equation}
where we chose $\Delta = 1.5,\ \Delta^\prime = 0.5$ in the numerical simulation. Here we consider magnetization transport, i.e., the seed operator is magnetization density wave like Eq.\ (\ref{Wave}), where our local operator is $\hat{A}_n = s^z_n$. The Heisenberg chain is integrable if only nearest-neighbor interactions are present, meaning $\Delta' = 0$. In this case, energy and spin transport are ballistic for $|\Delta|<1$, whereas spin transport becomes diffusive for $|\Delta|>1$ \cite{bertini2021finite}. However, introducing next nearest-neighbor interaction breaks integrability.

\begin{figure}[t]
\includegraphics[width=1\linewidth]{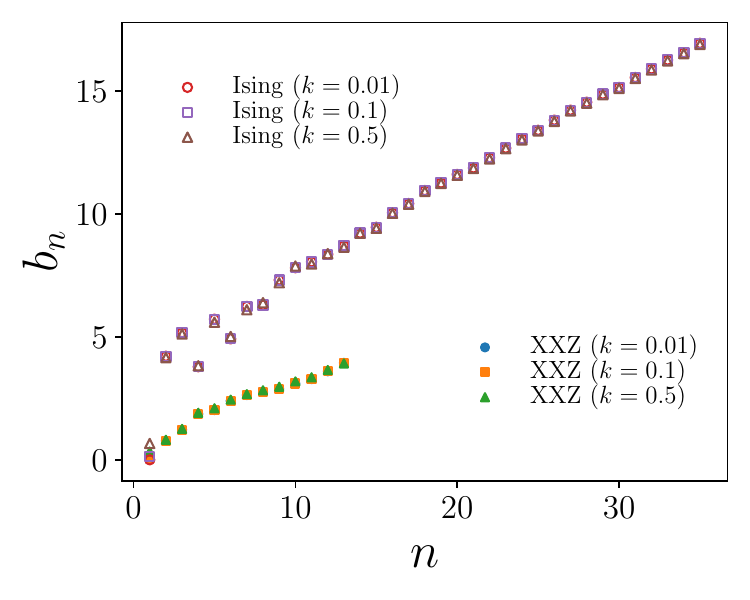} 
    \caption{The Lanczos coefficients $b_n$ for both a Heisenberg XXZ model, as well as a tilted-field Ising model denoted as $I$ are plotted against $n$ for different wave numbers. The wave numbers used here are $k_1 = 0.01$, $k_2=0.1$ and $k_3 = 0.5$. For the Heisenberg chain $13$ coefficients were calculated, while $35$ were calculated for the Ising chain.}
    \label{fig5}
\end{figure}

\begin{figure}[t]
    \includegraphics[width=1\linewidth]{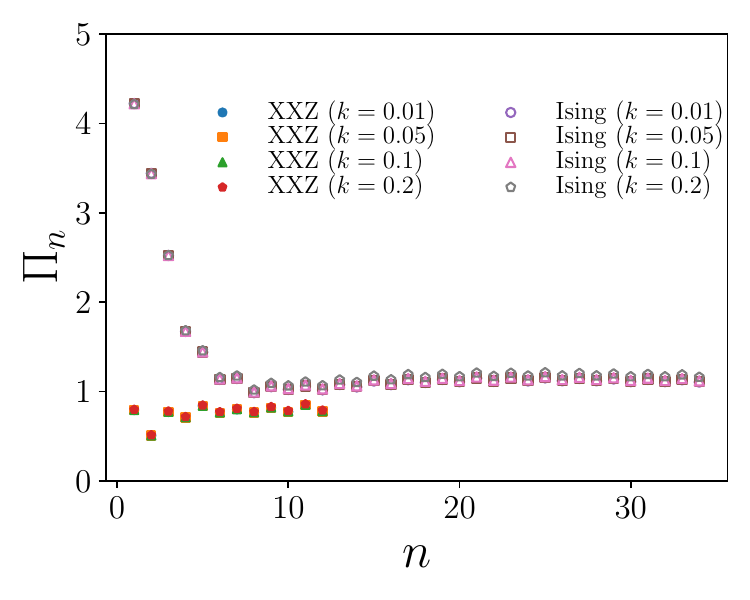} 
    \caption{The partial product $\Pi_n$ are plotted against $n$. Again for both the Heisenberg XXZ model and the titled field Ising model, the wave numbers were chosen to be $k_1 = 0.01$, $k_2 = 0.05$, $k_3 = 0.01$ and $k_4 = 0.02$.}
    \label{fig6}
\end{figure}

\begin{figure}[t]
    \includegraphics[width=1\linewidth]{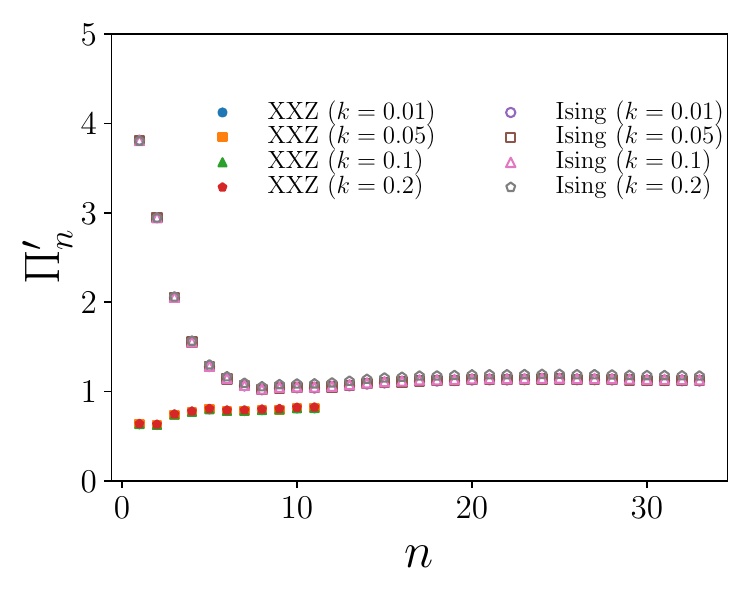} 
    \caption{Here, the partial product $\Pi'_n$ are plotted against $n$, where the geometric mean between two neighboring points in the sequence was taken. The wave numbers were chosen to be $k_1 = 0.01$, $k_2 = 0.05$, $k_3 = 0.01$ and $k_4 = 0.02$ for both the tilted field Ising and Heisenberg XXZ model.}
    \label{fig7}
\end{figure}
It shall first be noted that calculating Lanczos coefficients for a Heisenberg XXZ chain is more computationally costly than for the tilted-field Ising model. This in turn means that we only calculated the coefficients up to the number $13$ for the XXZ chain, while $35$ were calculated for the tilted-field Ising chain, as can seen in Fig. \ref{fig5}. The different wave numbers $k$ only lead to slight deviations that are barely visible. In the beginning, meaning for small $n$, the Lanczos coefficients deviate somewhat from a strictly linear growth but smoothly grow as $n$ increases. They then follow a linear growth pattern in line with the hypothesis suggested in \cite{parker2019universal} and \cite{heveling2022numerically} for larger values of $n$. Apart from the Lanczos coefficients, we also plotted the partial products $\Pi_n$ versus $n$ to receive insight into their convergence. As demonstrated in Fig.\ \ref{fig6}, after the initial data points the partial products seem to be convergent to a fixed finite value for all cases examined, although the pattern is not completely smooth, since there is a pattern of staggering around the constant trend lines. Therefore, we define
\begin{equation}\label{pin}
\Pi'_n \coloneq \sqrt{ \Pi_n \Pi_{n+1}},
\end{equation}
where geometric mean of two neighboring data points is taken in line with \cite{joslin1986calculation} with the aim of smoothing this sequence in $n$. These data points are plotted against $n$ in Fig.\ \ref{fig7}\\
After having examined the convergence for these physical systems, we arrive at the conclusion that pathological behavior may not occur in the investigated examples. Therefore, the claim that a finite area under the current autocorrelation function implies diffusive behavior can still be expected to hold within physical systems.

\section{Conclusion}

Central to this work are concepts such as density waves, currents and their respective autocorrelation functions. The relations between were examined in detail. Furthermore, the implications of an operator following the diffusion equation for the associated correlation function were discussed, namely an exponential decay. The dynamics of a density wave were cast into the form of a Volterra equation.  For the case of a wave number $k=0$ it was shown that the current is identical to the memory function. However, the transition from a vanishing to a small wave number is nontrivial. Therefore, a counterexample to the claim that a current has a finite area under its correlation function with respect to time was constructed, violating the Einstein relations.\\
Following this, two equivalent formalisms for examining the time evolution of the autocorrelation for a given operator, were introduced. On the one hand there is the Lanczos algorithm with its intuitive interpretation as a semi-infinite hopping chain. Secondly, the Mori formalism was introduced to cast the dynamics into a nested set of Volterra equations, where each such equation governs the memory dynamics of the previous one. Within the Mori memory formalism the precise difference between the current correlation function and the memory function can be examined. This difference is the Liouvillian according to which the time evolution is governed. In the case of the current it is the full Liouvillian, whereas for the memory function it is a projected Liouvillian equivalent to setting the first Lanczos coefficient $b_1(k) = 0$. Precisely this difference allows us to construct a counterexample that violates the Einstein relations.\\
An additional necessary condition for diffusion was established, by the construction of an infinite product involving Lanczos coefficients. Convergence of this product implies that the area under the memory function is finite, suggesting Markovian memory dynamics, leading to an exponentially decaying density profile. For the proposed counterexample, as was demonstrated numerically, this product does not converge even for small wave numbers. For an exponential decay, which represents the most clear case of diffusion, conversion becomes apparent quickly. Additionally, real physical models in the diffusive regime, namely the tilted field Ising model and the Heisenberg chain with next nearest neighbor interaction, were examined. As expected, the product in these cases also converges. Since these convergences become apparent even for the small number of coefficients calculated, the additional necessary condition is justified as an indication for presence of diffusive behavior.
The main question that remains is whether or not there are physical reasons that rule out counterexamples such as the one we have shown. It was constructed purely to serve as a counterexample for diffusion on the level correlation functions. It could, however, turn out that if one starts out with a given observable and a Hamiltonian that these scenarios are ruled out by some line of reasoning. This seems plausible given that we do not know of any physical counterexample to the claim we investigated in this paper.

\section*{Acknowledgments}

We thank Mariel Kampa, Markus Kraft and Merlin F\"ullgraf for fruitful discussions.  This work has been funded by the Deutsche
Forschungsgemeinschaft (DFG), under Grant No. 531128043, as well as under Grant
No.\ 397107022, No.\ 397067869, and No.\ 397082825 within the DFG Research
Unit FOR 2692, under Grant No.\ 355031190.

\bibliography{sample.bib}
\bibliographystyle{apsrev4-1_titles.bst}

\begin{appendix}

\section*{Appendix: Lanczos coefficients from a correlation function}\label{AppB:A}
\setcounter{equation}{0}
\renewcommand{\theequation}{A\arabic{equation}}

In Section \ref{Lanczos2} the Lanczos coefficients were computed by directly using of correlation functions. In this appendix we want to review how these results were computed, along the lines of \cite{joslin1986calculation}. To follow along these lines one must first compute moments of a given correlation function. These will be referred to as $M_{2n}$. Here only even moments contribute, and all odd moments of a given correlation function are zero since at infinite temperature correlation functions are even functions with respect to time. These will therefore be calculated according to
\begin{equation}
    M_{2n} \coloneq \frac{d^{2n}}{dt^{2n}} C(t) |_{t=0}
\end{equation}
We will now use these to define an auxiliary quantity, namely $c_n$, according to
\begin{equation}
    c_n \coloneq \left| \frac{M_{2n}}{M_0} \right|
\end{equation}
With this quantity it is possible to create matrices, whose determinants are of interest. First we define $B_0 = B_1 \coloneq 1$ and for $n \geq 2$ we calculate the $B_n$'s according to
\begin{equation}
    B_n \coloneq \begin{vmatrix}
        1 & c_1 & \cdots & c_{n-1} \\
        c_1 & c_2 & \cdots & c_n \\
        \vdots & \vdots & \ddots & \vdots \\
        c_n & c_{n+1} & \cdots & c_{2n-1}
    \end{vmatrix}.
\end{equation}
Furthermore, we will also define $C_n$'s, where $C_0 \coloneq 1$ and for $n \geq 1$ we define
\begin{equation}
    C_n \coloneq \begin{vmatrix}
        c_1 & c_2 & \cdots & c_n \\
        c_2 & c_3 & \cdots & c_{n+1} \\
        \vdots & \vdots & \ddots & \vdots \\
        c_n & c_{n+1} & \cdots & c_{2n-1}
    \end{vmatrix}.
\end{equation}
This now allows us to calculate the Lanczos coefficients, where the calculation differs for the even and the odd case,
\begin{align}
    b_{2n-1} = \sqrt{\frac{B_{n-1} C_n}{B_n C_{n-1}}} \\
    b_{2n} = \sqrt{\frac{B_{n+1} C_{n-1}}{B_n C_n}}.
\end{align}
\end{appendix}

\end{document}